\documentclass[useAMS,usenatbib]{mn2e}

\usepackage{txfonts,graphicx,amssymb,subfigure}
\usepackage[normalem]{ulem}
\usepackage{color}

\title{Deciphering solar turbulence from sunspots records}

\author[F.~Plunian et al.]
       {F.~Plunian$^{1}$, G.~ R.~Sarson$^{2}$, R.~Stepanov$^{3}$\\
$^1$ LGIT, UJF, CNRS, B.P. 53, 38041 Grenoble Cedex 9, France \\
$^2$ School of Mathematics and Statistics, Newcastle University, Newcastle NE1 7RU, UK \\
$^3$ Institute of Continuous Media Mechanics, Korolyov 1, 614013 Perm, Russia}

\date{Accepted 2009 .... Received 2009 ....; in original form 2009}

\pagerange{\pageref{firstpage}--\pageref{lastpage}}
\pubyear{2009}
\begin{document}
\maketitle

\label{firstpage}

\begin{abstract}
It is generally believed that sunspots are the emergent part of magnetic flux tubes in the solar interior.
These tubes are created at the base of the convection zone and rise to the surface due to their magnetic buoyancy.
The motion of plasma in the convection zone being highly turbulent,
the surface manifestation of sunspots may retain the signature of this turbulence, including its intermittency.
From direct observations of sunspots,
and indirect observations of the concentration of cosmogenic isotopes
$^{14}$C in tree rings or $^{10}$Be in polar ice,
power spectral densities in frequency are plotted.
Two different frequency scalings emerge, depending on whether the
Sun is quiescent or active.
%magnetic activity is maximum or minimum.
From direct observations we can also calculate scaling exponents.
These testify to a strong intermittency,
comparable with that observed in the solar wind.
\end{abstract}


\section{Introduction}
\begin{figure} \centering {
    \includegraphics[width=0.4\textwidth]{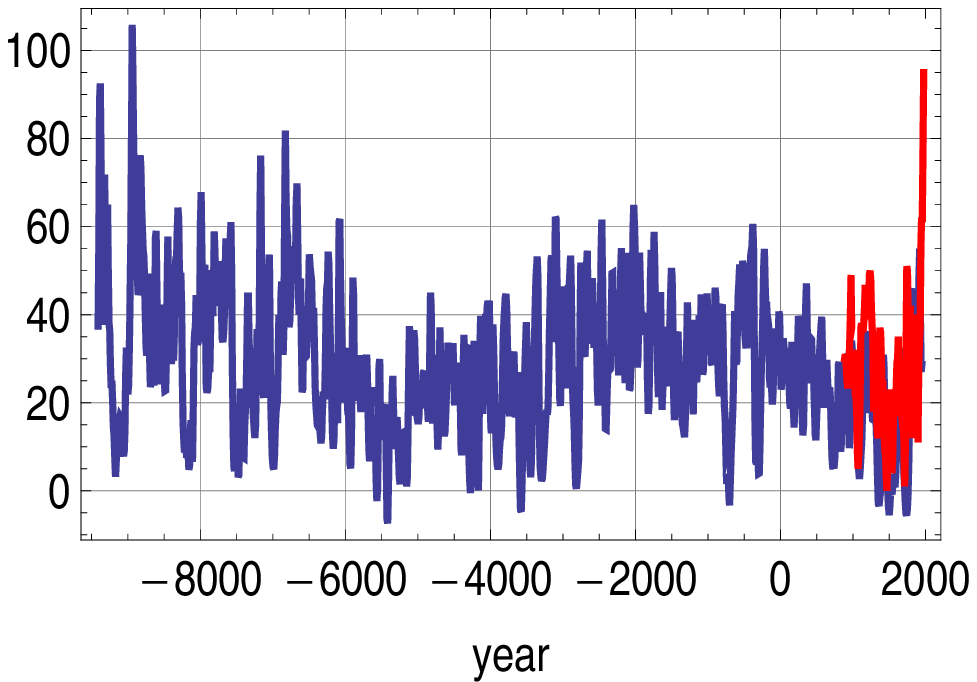}
    \includegraphics[width=0.4\textwidth]{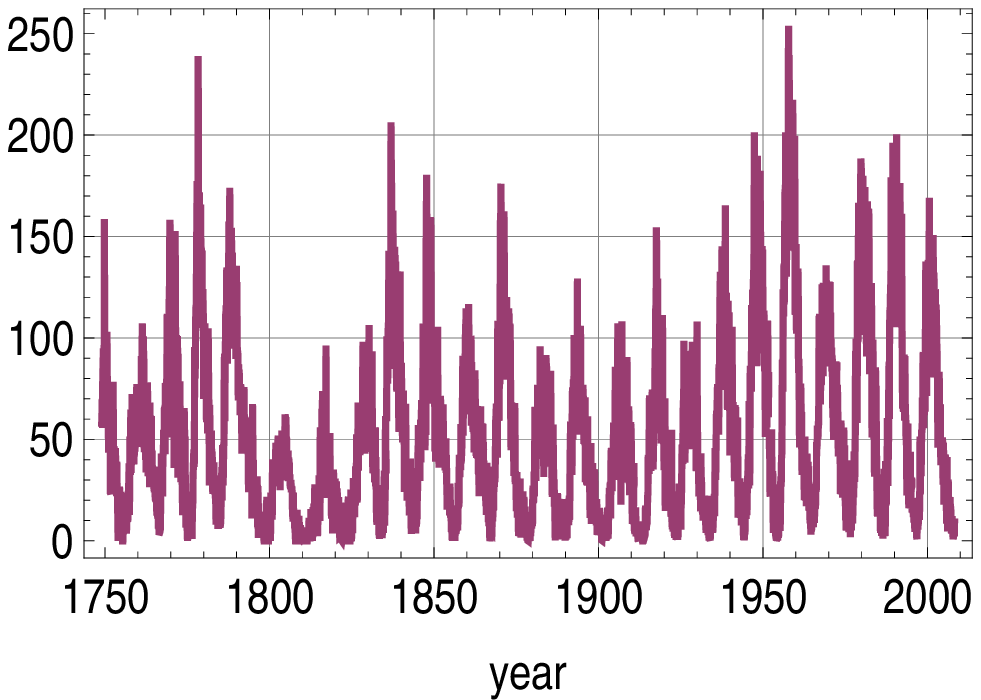}
    \includegraphics[width=0.4\textwidth]{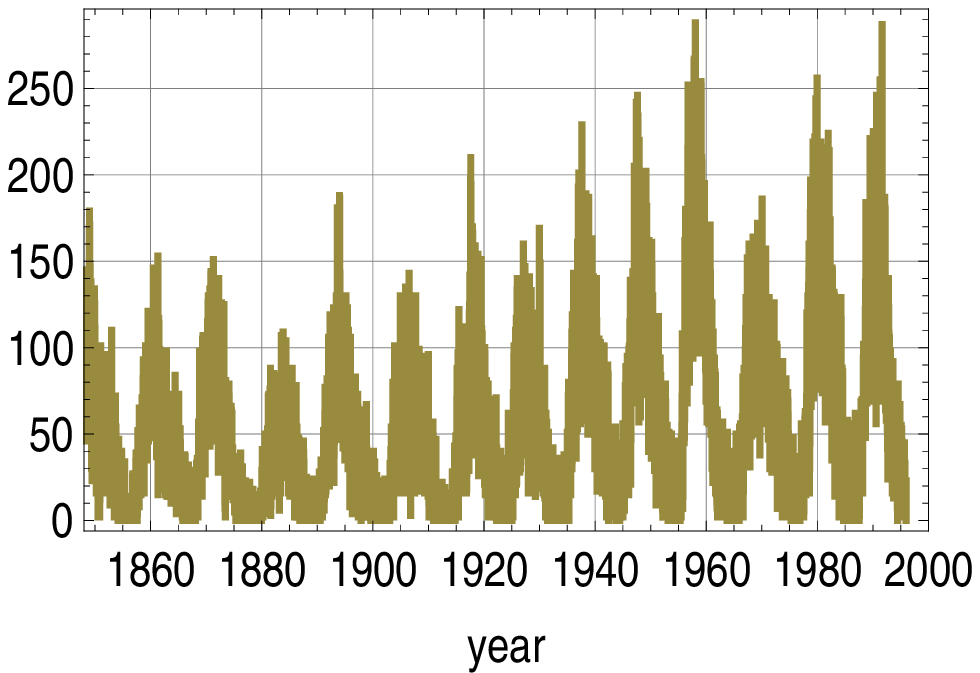}
    \includegraphics[width=0.4\textwidth]{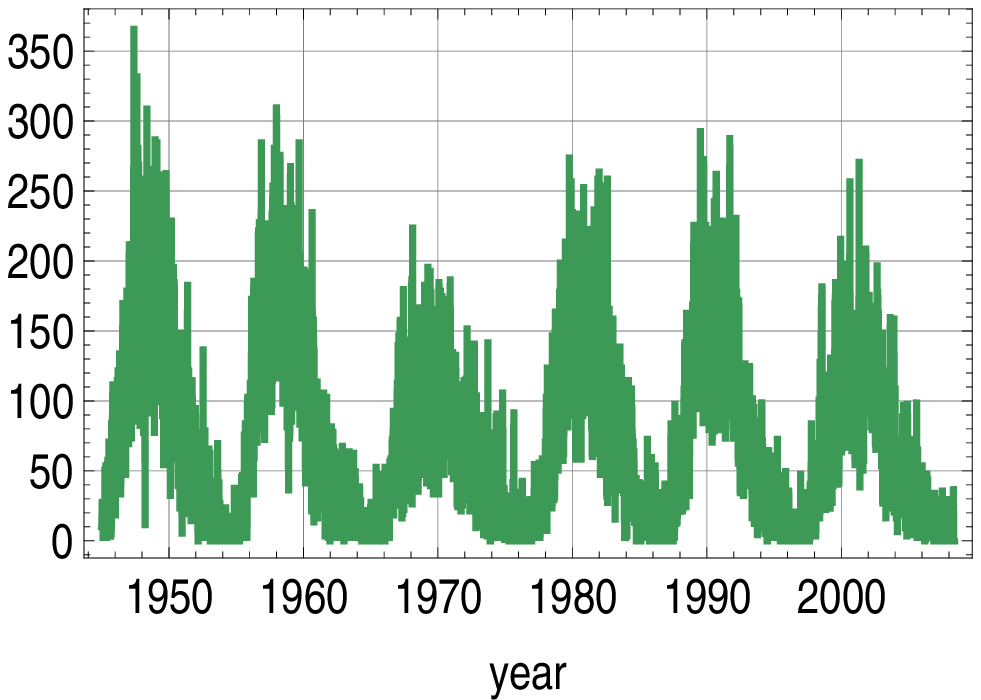}}
    \caption{Data sets of 10 year-averaged SSN from $^{14}$C (top blue) and $^{10}$Be (top red), International SSN monthly averaged M (second),
    American daily group SSN G (third), and American daily SSN D (bottom).}
    \label{data}
\end{figure}

Sunspots observed at the surface of the convection zone of the Sun
are usually understood as the manifestation of solar magnetic activity.
Naked-eye and telescope observations of sunspots are available from AD 1610,
providing reliable records of sunspot numbers (SSN).
Several sets of data exist, varying in how they have been sampled,
averaged (daily or monthly), whether they concern sunspots or sunspots groups,
and on the scientific societies who have compiled the records.
Here we consider the American daily SSN (D) \footnote{ftp://ftp.ngdc.noaa.gov/STP/SOLAR\_DATA/SUNSPOT\_NUMBERS/\\AMERICAN\_NUMBERS/RADAILY.PLT},
the American daily group SSN (G) \footnote{ftp://ftp.ngdc.noaa.gov/STP/SOLAR\_DATA/SUNSPOT\_NUMBERS/\\GROUP\_SUNSPOT\_NUMBERS/dailyrg.dat},
and the International monthly averaged SSN (M) \footnote{http://solarscience.msfc.nasa.gov/greenwch/spot\_num.txt} (figure \ref{data}).
In addition SSN at earlier times have been reconstructed from proxies,
based on the concentration of cosmogenic isotopes $^{14}$C in tree rings \footnote{ftp://ftp.ncdc.noaa.gov/pub/data/paleo/climate\_forcing/solar\_variability/\\solanki2004-ssn.txt}
or $^{10}$Be in ice core bubbles.
The production rate of such isotopes increases with the cosmic-ray flux,
which is higher when the solar magnetic activity is low.
Plotting the SSN versus time
reveals a cycle of about 11 years known as the Schwabe cycle.
This cyclical solar magnetic activity is sufficiently robust
to be detected in $^{10}$Be concentration records,
even during some long periods with almost no visible sunspots,
like the Maunder minimum (1645-1715) \citep{Beer98}.
Analyzing long time series of $^{14}$C and $^{10}$Be,
it has been shown that the solar activity of the last 70 years
has been exceptionally high \citep{Usoskin03,Solanki04},
and that a decline is expected within the next two or three cycles \citep{Abreu08}.

The occurrence of sunspots is of course an important diagnostic that must be reproduced by any solar dynamo model.
The fact that it is irregular (in spite of the Schwabe cycle)
reflects the complexity inherent in the nonlinear coupling
between the turbulent flow and the magnetic field in the solar convection zone \citep{Browning06}.
With the long time series of direct SSN observations and
$^{14}$C and $^{10}$Be data available,
it is tempting to calculate the corresponding frequency spectra,
to infer some signature of the underlying turbulence in the convection zone.
Similar attempts have been made for the Earth \citep{Courtillot88, Constable05, Sakuraba07};
the frequency spectrum of the geomagnetic dipole moment obtained from paleomagnetic data
is consistent with an inertial range of $f^{-5/3}$ scaling,
and a dissipation range of $f^{-11/3}$ as expected in magnetohydrodynamic turbulence \citep{Alemany00}.
For the Sun, analysis of International daily SSN have led to a $f^{-2/3}$ scaling \citep{Morfill91,Lawrence95},
and this has been attributed to some sequential sampling of the field upon arrival at the
photosphere on top of a Kolmogorov spatial scaling due to the underlying turbulence.
Here we extend this analysis to the other SSN records mentioned above and test how robust the $f^{-2/3}$ scaling is, in particular during minima and maxima of sunspot activity. In addition, for time scales smaller than 2 years, the stochastic character of the SSN records
suggests strong intermittency \citep{Lawrence95}, as opposed to a low-dimensional chaotic interpretation. We shall characterize this intermittency by calculating the corresponding scaling exponents.
%From the SSN records, in addition to the spectra we can also characterize
%the intermittency, provided that the sampling is sufficiently high ($\le$ 1 month)
%and the records cover a sufficiently long time.

\section{Wavelet spectra}

\begin{figure}\centering{
    \includegraphics[width=0.46\textwidth]{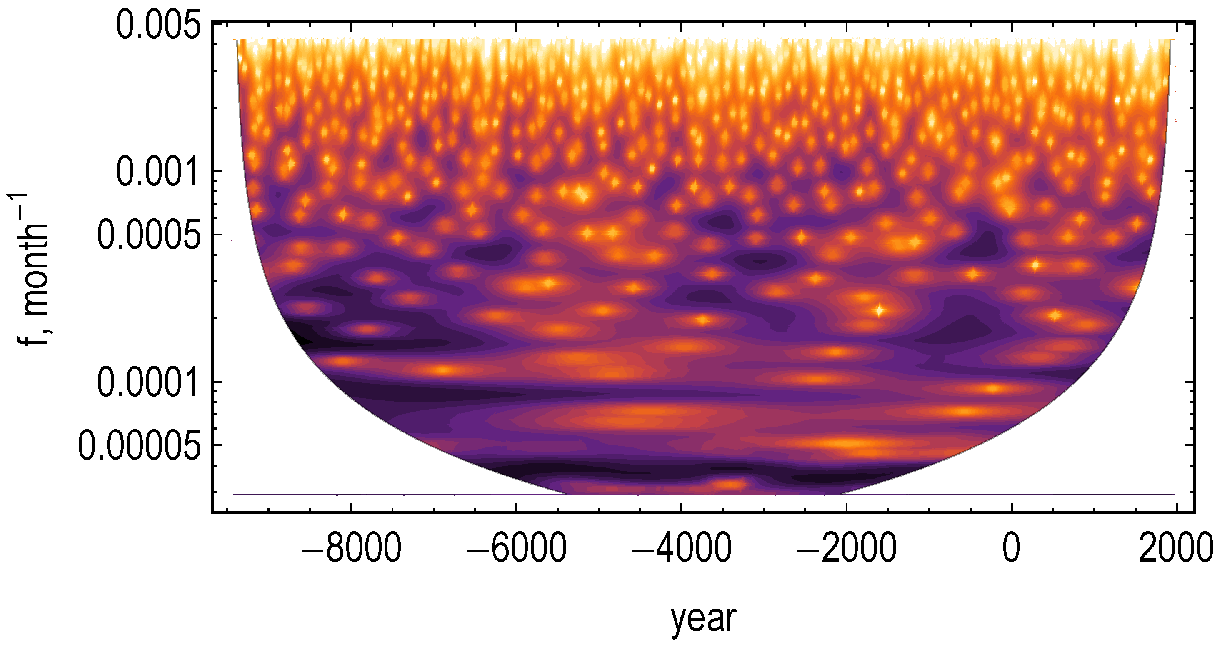}
    \includegraphics[width=0.46\textwidth]{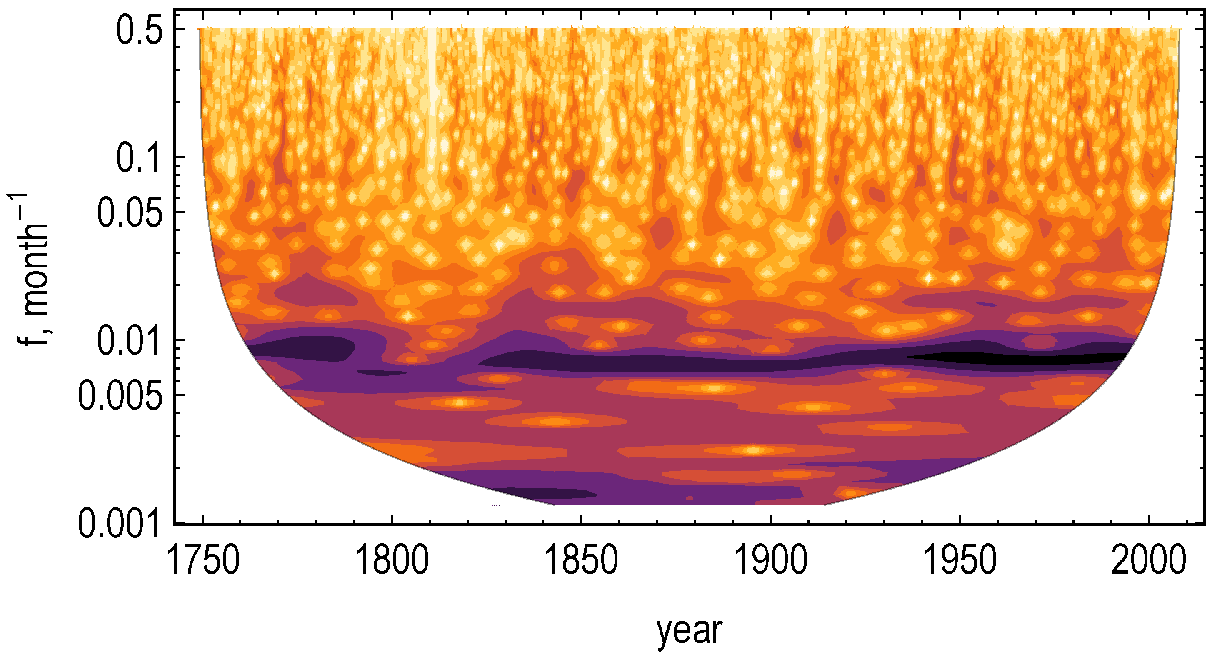}
    \includegraphics[width=0.46\textwidth]{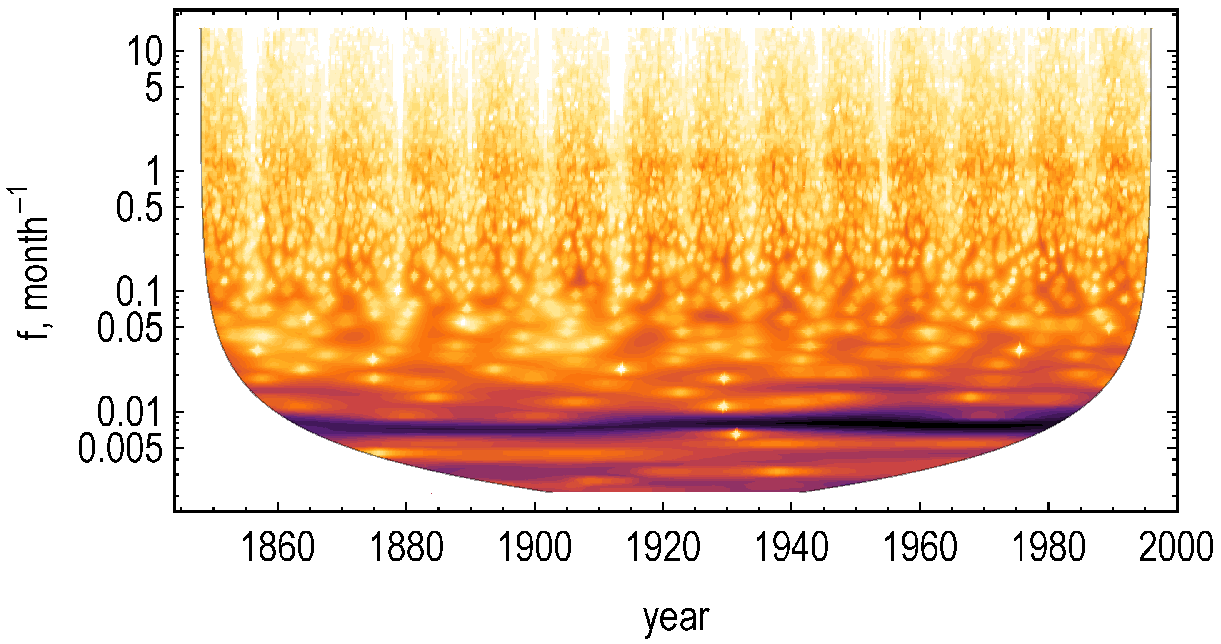}
    \includegraphics[width=0.46\textwidth]{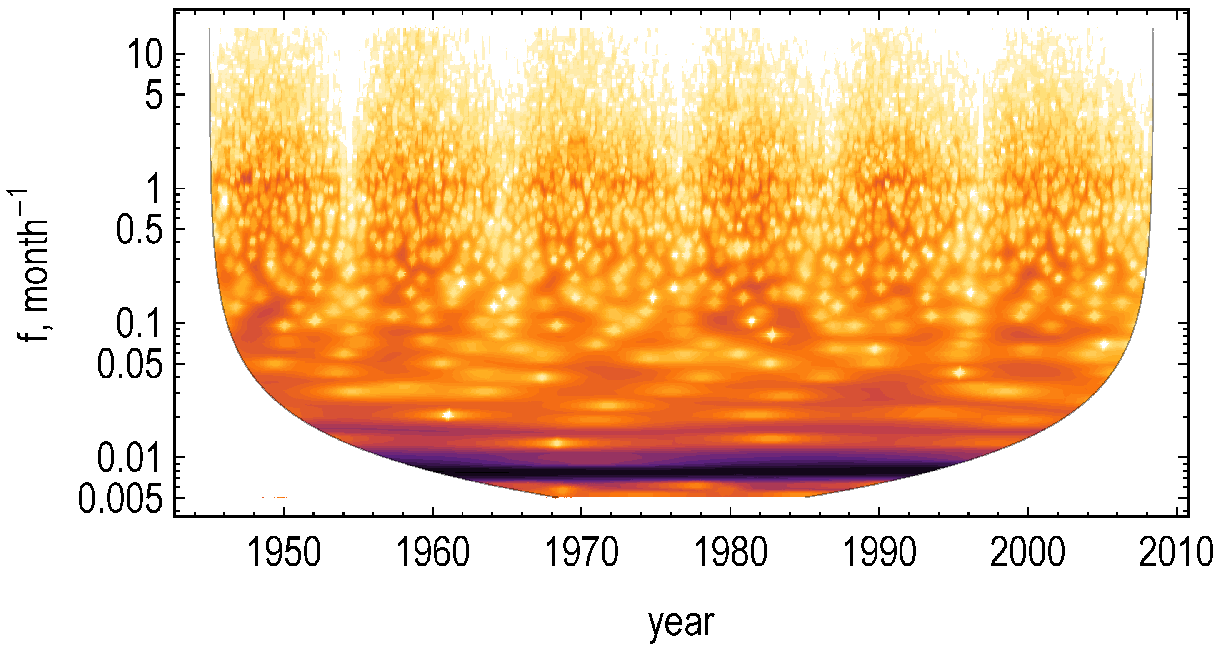}}
    \caption{Wavelet coefficients (logarithm of absolute value) of $^{14}$C (top), M (second), G (third), and D (bottom).}
    \label{wavelet}
\end{figure}

In order to filter out the noise, we use a wavelet decomposition of the signal.
In figure \ref{wavelet}, the wavelet coefficients are plotted versus time (in years) and frequency (in month$^{-1}$).
The light (resp.\ dark) colors correspond to low (resp.\ high) values of these coefficients.
For a given year, the curve giving the wavelet coefficient versus frequency corresponds to a power spectral density of the signal.
The dark horizontal stripe for $f\sim 0.01$ month$^{-1}$
corresponds to the Schwabe cycle.
It is not visible in the $^{14}$C data set due to the coarse sampling of this data (averaged over 10 years).
On the other hand a clear dark stripe is visible for $f\sim 3.5 \times 10^{-5}$ month$^{-1}$, corresponding to a $\sim$2400 years cycle.
In the figure for M data the dark stripe of the Schwabe cycle almost disappears during the Dalton minimum (1790-1820) \citep{Frick97}.
In the G and D data figures we identify another horizontal stripe at $f\sim 1$ month$^{-1}$,
corresponding to the solar mean rotation rate. This indicates that the latitudinal repartition of sunspots
is not homogeneous.
Finally the vertical light stripes correspond to minima of magnetic activity.

\begin{figure}\centering{
    \includegraphics[width=0.4\textwidth]{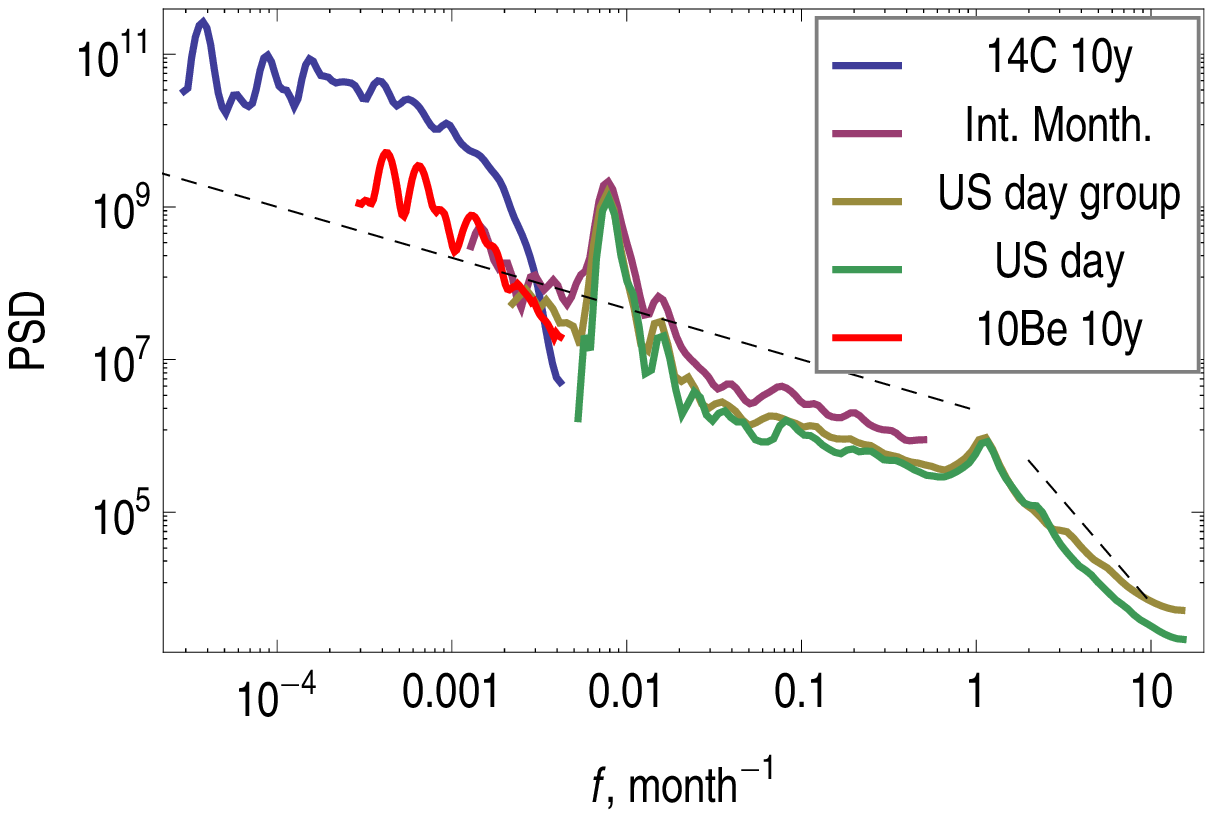}
    \includegraphics[width=0.4\textwidth]{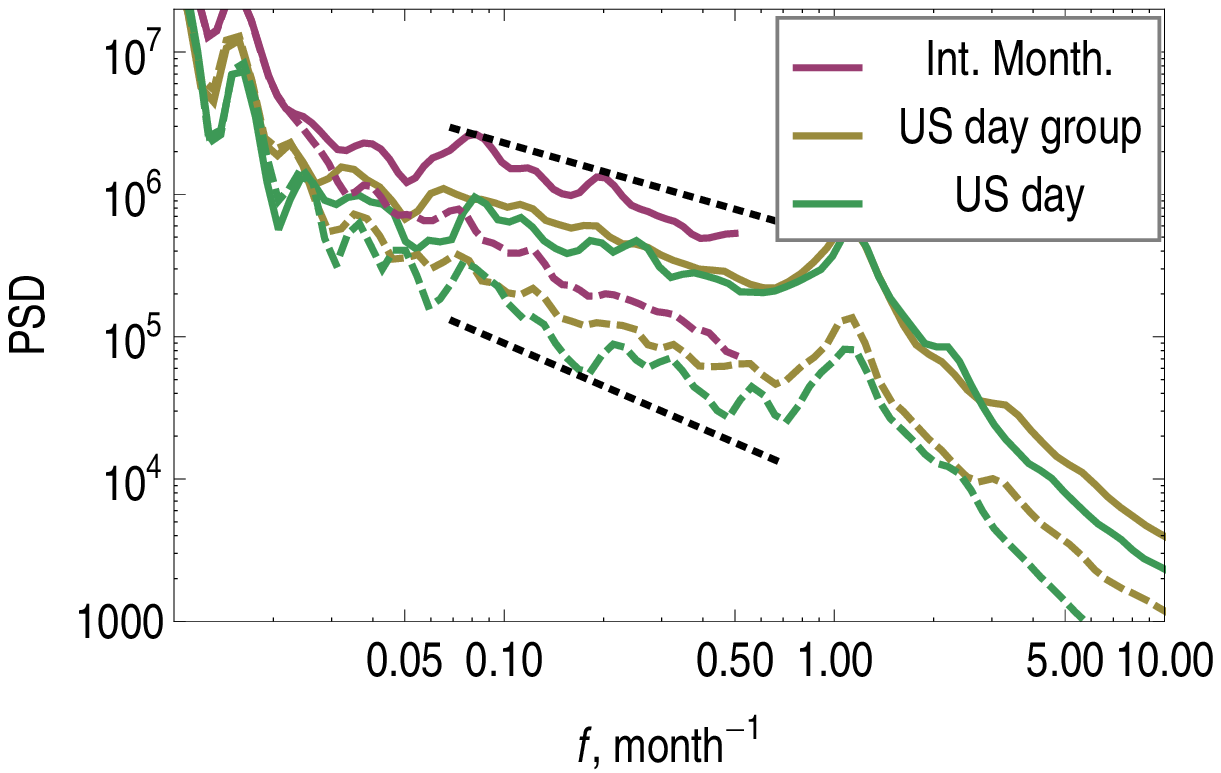}}
    \caption{Wavelet spectral density versus frequency, on average (top), and for magnetic activity minima (bottom, dashed curves) and maxima (bottom, solid curves). The dashed straight lines correspond to $-2/3$ and $-8/3$ slopes (top), and  $-2/3$ and $-1$ slopes (bottom).}
    \label{spectra}
\end{figure}

The time average of the wavelet coefficients are shown in figure \ref{spectra} (top) for the five data sets.
As mentioned earlier the two peaks around $0.01$ month$^{-1}$ and $1$ month$^{-1}$ correspond to the Schwabe cycle and solar mean rotation rate.
Between them
the three spectra of the directly observed SSN (M, G, D) present a common scaling in $f^{-2/3}$.
The other data sets $^{14}$C and $^{10}$Be are compatible with a $f^{-2/3}$ scaling as well, although
the $^{14}$C PSD is overestimated by roughly a factor 10, probably due to the proxy used in the reconstruction
of the SSN from the $^{14}$C data.

\begin{figure} \centering{
    \includegraphics[width=0.4\textwidth]{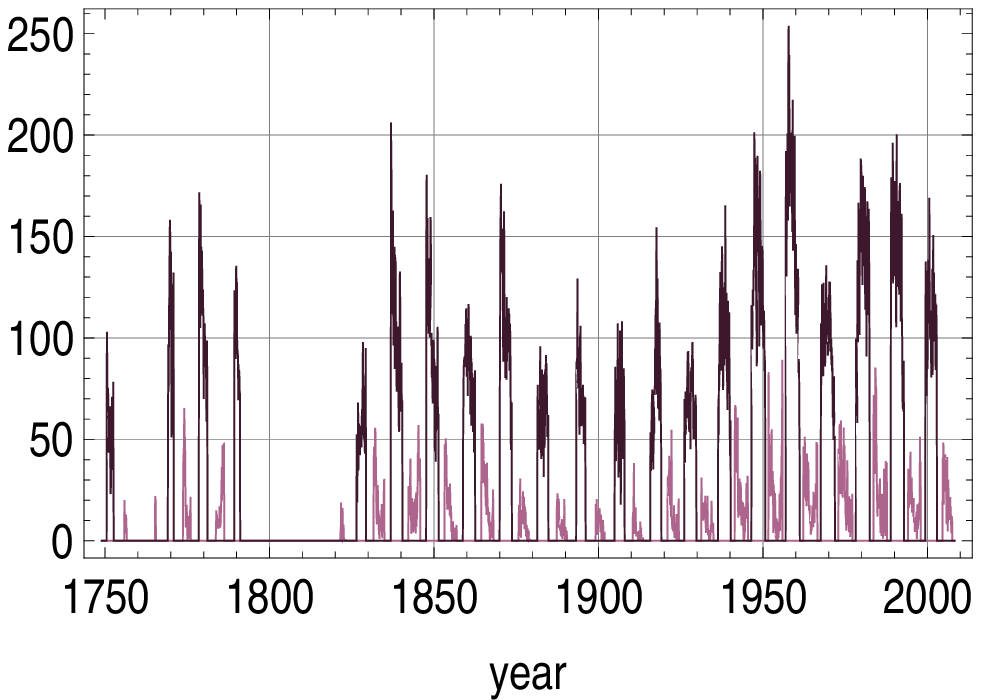}
    \includegraphics[width=0.4\textwidth]{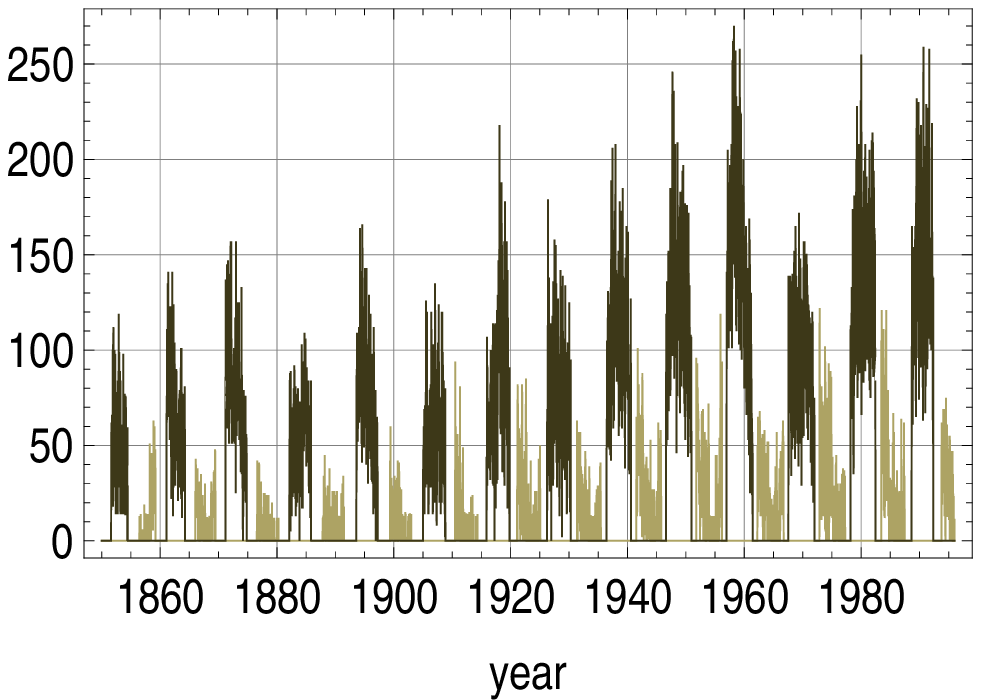}
    \includegraphics[width=0.4\textwidth]{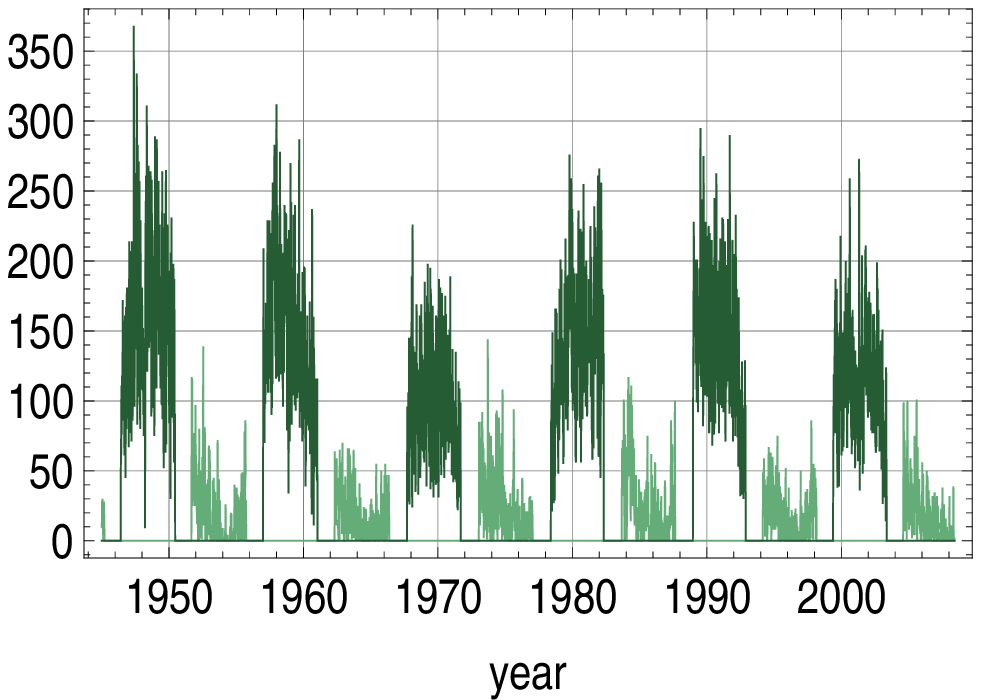}
    }
%\caption{Same curves as in figure \ref{data} (three last ones) with sets of maximum (dark) and minimum (light) magnetic activity.}
\caption{Same data as in figure \ref{data} (the last three panels) but subdivided into sets of maximum (dark) and minimum (light) magnetic activity.}
    \label{splitspectra}
\end{figure}

For the three data sets (M), (G) and (D), instead of averaging on all times, we now average on periods corresponding to either maximum or minimum magnetic activity as shown in figure \ref{splitspectra}. {For the maximum (resp.\ minimum) activity subset, the excluded data is that centred around the times of the Schwabe minima (resp.\ maxima)}  The corresponding spectral densities are plotted in figure \ref{spectra} (bottom).
The slopes for minima are systematically steeper than those for maxima, indicating two different regimes.
{
To estimate these slopes we vary both the range of frequency $\left[ f_{\min}, f_{\max}\right]$ on which they are calculated, and the way the data sets are split into subsets of maximum and minimum activity. For the former we take $f_{\max}=0.7$ month$^{-1}$ for the data sets (G) and (D) in order to escape from the influence of the peak $f=1$ month$^{-1}$, and $f_{\max}=0.5$ month$^{-1}$ for the data set (M). When changing $f_{\min}$ the slopes change. We vary $f_{\min}$ such that the ratio $f_{\max}/ f_{\min}$ is about 10, and the standard deviation of the slope remains 10 \% or less of its average value. This leads to $f_{\min}\in\left[0.05, 0.07\right]$ for (G), $f_{\min}\in\left[0.06, 0.08\right]$ for (D), and $f_{\min}\in\left[0.03, 0.05\right]$ for (M).
In addition we consider at least three different degrees of splitting for each data set,
this splitting degree being related to the time length of the subsets of maximum and minimum activity.
The choices of this splitting degree are made such that both subsets are long enough to provide good statistics, but remain separated by sufficient time lags that the SSN values falling into each subset do not overlap too much.
Then for both the maximum and minimum subsets, we calculate the mean slope and the standard deviation obtained when varying both the frequency range and the degree of splitting.
The corresponding slope estimates are given in table \ref{slopes}.
They are consistent with power spectra in $f^{-2/3}$ and $f^{-1}$. The corresponding dashed lines are plotted in figure \ref{spectra} to guide the eye.
%ngrs: the following added, to emphasise the point.
The standard deviations are small, showing that these slopes are robust with respect to the details of our analysis.
%ngrs: is it also worth noting the typical magnitude of the standard errors from the formal regressions, as follows?
The formal standard errors from each of the individual regressions (for specific frequency ranges and degrees of splitting) are of comparable magnitude.
\begin{table}
\begin{tabular}{c|ccc}
  Activity & (M) & (G) & (D)  \\
  \hline
  Max. & $-0.61 \pm 0.05$ &$-0.69 \pm 0.05$ & $-0.63 \pm 0.05$ \\
  Min. & $-1.03 \pm 0.03$ &$-0.94 \pm 0.07$ & $-0.98 \pm 0.02$ \\
\end{tabular}
\caption{Slope estimates for the PSD curves plotted in figure \ref{spectra} (bottom).
They correspond to average values plus standard deviation errors when varying both the frequency range and the degree of splitting of each data set (into the two subsets of maximum and minimum magnetic activity).}
\label{slopes}
\end{table}
}

As noted by Lawrence et al.\ (1995), the question of causality complicates the interpretation of such temporal data. The difference of spectral slopes between minima and maxima can be attributed to two different effects: a change of the underlying turbulence, affecting the spatial structure the magnetic field; or a change in the frequency of the sequential sampling of the magnetic field, as suggested by Lawrence et al.\ (1995).
Although the latter effect cannot be excluded, there is a simple argument in favour of the former.
It is generally accepted that the occurrence of sunspots at the photosphere
is due to the magnetic buoyancy force $\nabla B^2$, where $B$ is some magnetic induction intensity in the convection zone \citep{Tobias01}.  It is then tempting to interpret the two spectral slopes as the signatures of this buoyancy, assuming that the frequency of sunspot occurrence at the photosphere is proportional to this force.
Then the $f^{-2/3}$ and $f^{-1}$ SSN spectra would correspond to buoyancy spectra of $ k^{-2/3}$ and $k^{-1}$, where $k$ is the spatial wave number.
During maxima this implies a Kolmogorov magnetic energy spectrum of $k^{-5/3}$, compatible with inertia-driven turbulence in the convection zone. During minima it implies a magnetic energy spectrum of $k^{-2}$, compatible with turbulence dominated by the solar rotation \citep{Zhou95}.
In the transport scenario proposed by Tobias et al.\ (2001), the field which arises at the surface is the strongest part of a poloidal field generated by cyclonic turbulence in the convection zone.
Our interpretation then suggests two different regimes for such cyclonic turbulence, controlled by either inertia or rotation.

\section{Intermittency}

The stochastic nature of SSN occurrence for times scales smaller than 2 years has been shown by Lawrence et al.\ (1995), suggesting an intermittent turbulence. Here our goal is to quantify this intermittency for the three sets (M, G, D), calculating the corresponding scaling exponents.
For that we first calculate the associated generalized structure function (GSF)
and look for its scaling exponents, as usually done in turbulence.
We define the SSN increment by
\begin{equation}
\delta y(t,\tau) = S(t+\tau)-S(t) \; ,
\end{equation}
where $S(t)$ denotes the SSN at time $t$.
Assuming statistical stationarity in the frequency range of interest,
the $t$ dependence in $\delta y(t,\tau)$ can be dropped
and the GSF is then given by \citep{Nicol08}
\begin{equation}
S_m(\tau)= \left\langle \left|\delta y\right|^m\right\rangle=\int^{\infty}_{- \infty}\left|\delta y\right|^m P(\delta y, \tau) d(\delta y) \; ,
\end{equation}
where $P$ is the probability density function (PDF) of $\delta y$, the angle brackets $\left\langle \cdot\right\rangle$ denote time averaging,
and $m$ is a positive integer.

\begin{figure}
    \includegraphics[width=0.4\textwidth]{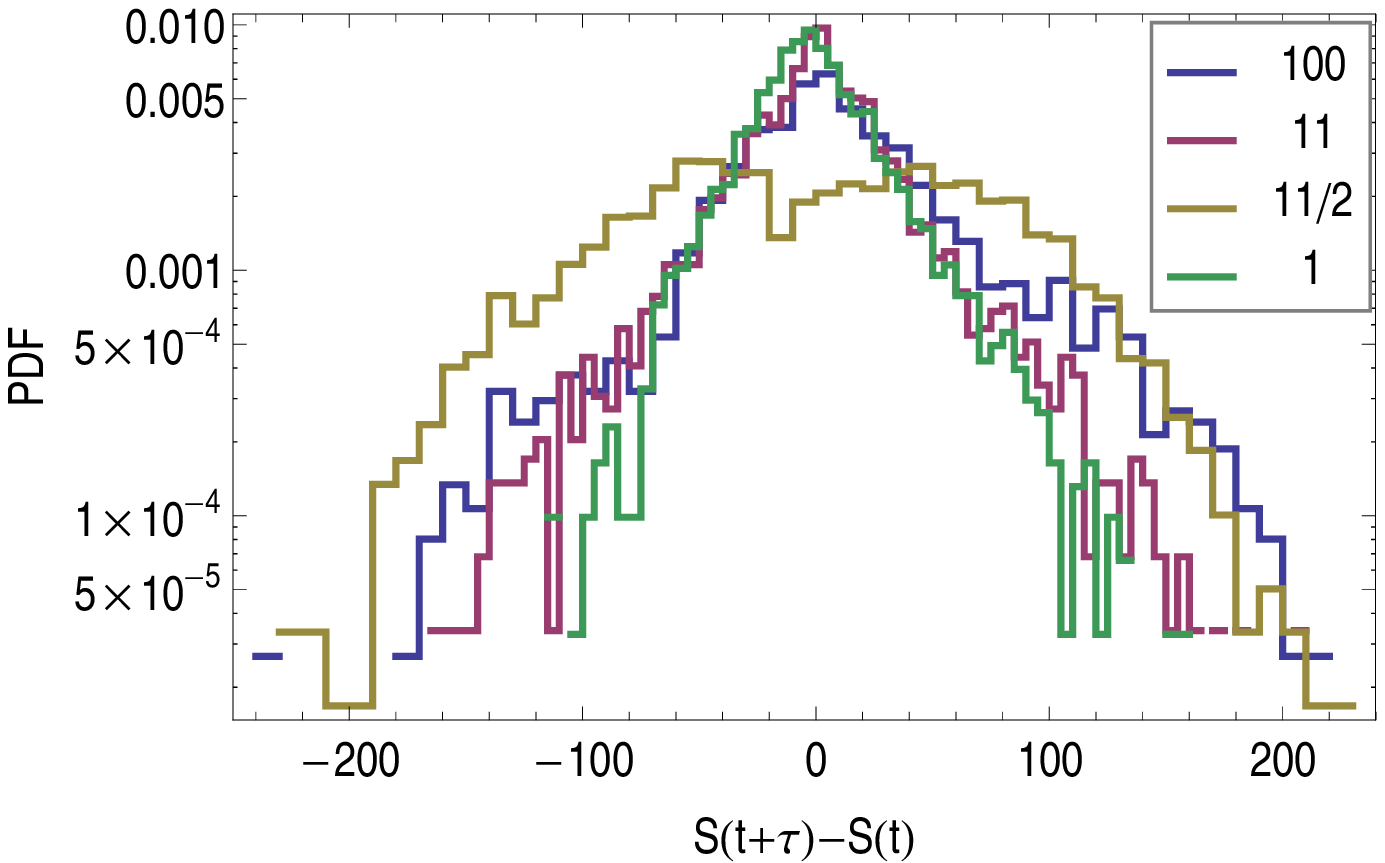}
    \includegraphics[width=0.4\textwidth]{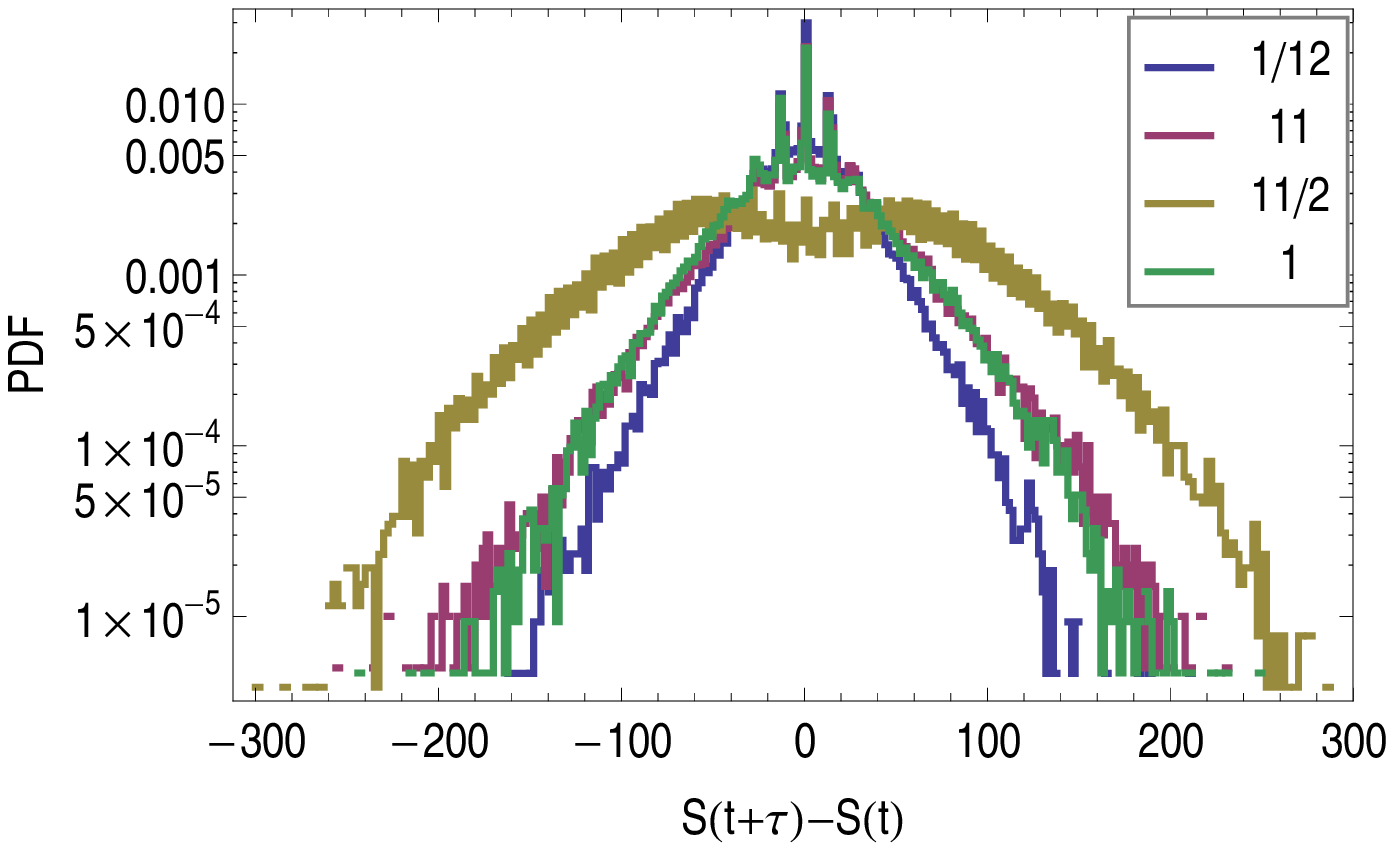}
    \includegraphics[width=0.4\textwidth]{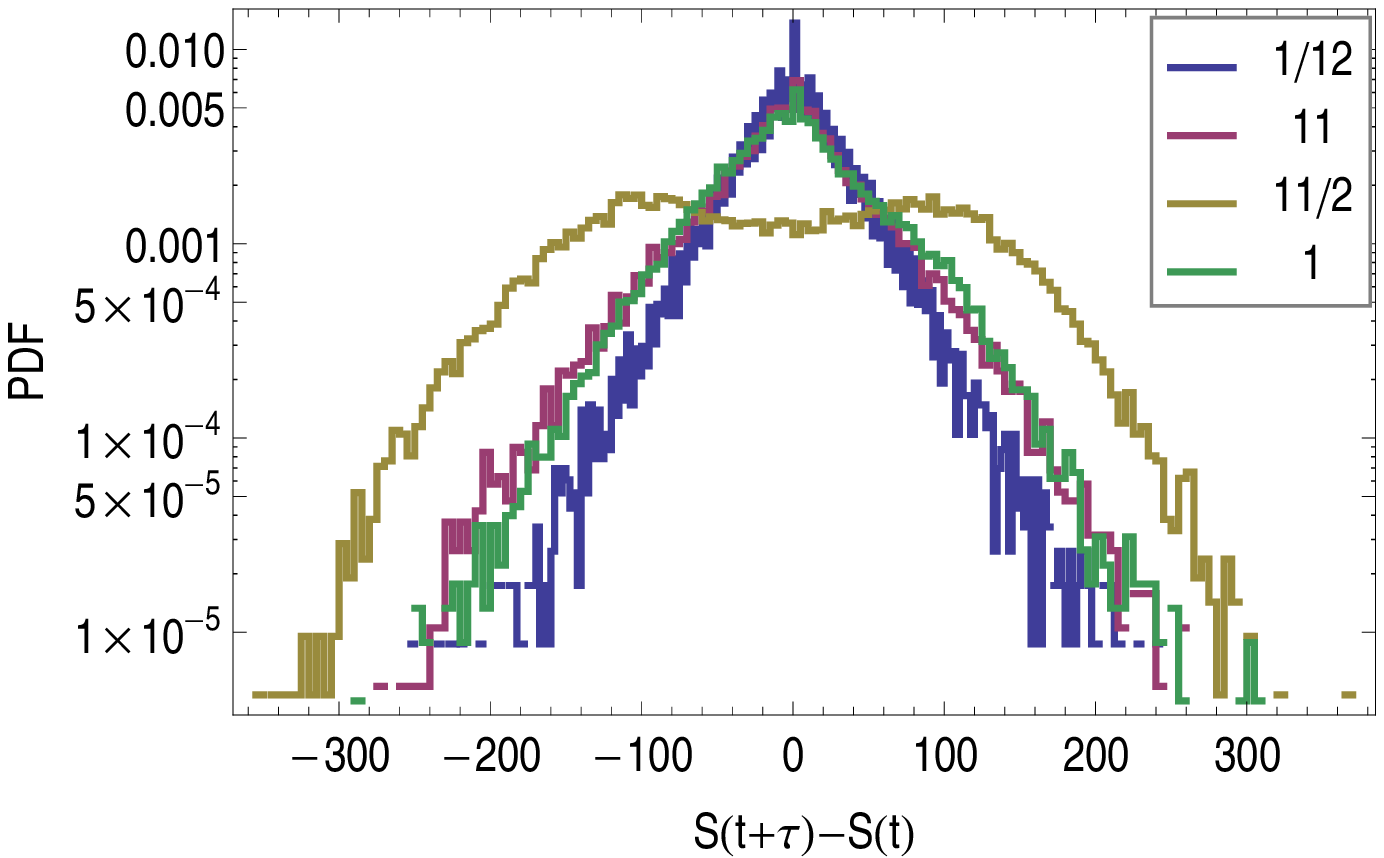}
    \caption{Probability density functions of $S(t+\tau)-S(t)$ for M (top), G (middle) and D (bottom). The labels indicate the value of $\tau$ in years.}
    \label{PDF}
\end{figure}

In figure \ref{PDF} PDFs for the three data sets (M, G, D) are plotted
for selected values of $\tau$.
%grs
%The PDFs of the $^{14}$C data are poorly described and we shall drop this data set in the rest of the study.
The PDFs of the $^{14}$C data are poorly defined, and so we drop this data set for the rest of the study.
The PDFs of the other data sets show peaks at $\tau=11/2$ years,
corresponding to the Schwabe cycle.
%grs
%For other values of $\tau$ they exhibit tails revealing a number of rare events higher than for a gaussian distribution, suggesting intermittency.
For other values of $\tau$ they exhibit tails containing a higher number of rare events than for a gaussian distribution, suggesting intermittency. Similar results were shown in Lawrence et al.\ (1995).

\begin{figure}
    \includegraphics[width=0.4\textwidth]{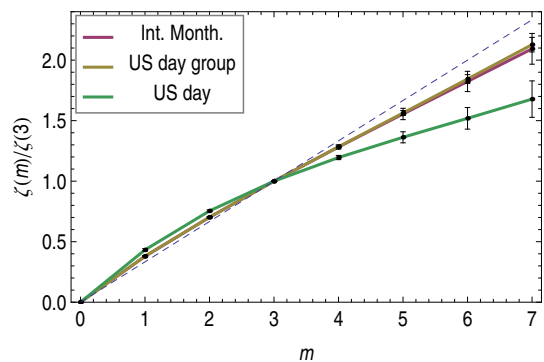}
    \caption{Scaling exponents $\zeta(m)/\zeta(3)$ plotted versus $m$ for the three sets of data M, G, D, and for 2 months $<\tau <$ 14 months. The dashed line corresponds to a Kolmogorov scaling $\zeta(m)/\zeta(3)=m/3$.}
    \label{scaling exponents}
\end{figure}

%grs
%To quantify this intermittency we first look whether the GSF obey a scaling law in the form
To quantify this intermittency we first check whether the GSF obey a scaling law in the form
\begin{equation}
	S_m(\tau) \sim \tau^{\zeta(m)}.
	\label{scaling}
\end{equation}
We find (not shown) that this is clearly the case.
In homogeneous and isotropic fully developed turbulence, intermittency corresponds to $\zeta(m) < m/3$.
The ratio $\zeta(m)/\zeta(3)$ is calculated, estimating the scaling power of $S_m(\tau) / S_m(3)$.
Plotting the ratio $\zeta(m)/\zeta(3)$ versus $m$ for the three data sets (figure \ref{scaling exponents})
%grs
%we see a clear departure of Kolmogorov straight line $\zeta(m)/\zeta(3)=m/3$, and then clear indications of intermittency.
we see a clear departure from the Kolmogorov straight line $\zeta(m)/\zeta(3)=m/3$, and clear indications of intermittency.
It is remarkable that the (D) set, which has the best sampling, leads to the largest intermittency.
It is also remarkable that the (M) and (G) sets lead to similar scaling exponents,
supporting the equivalence between averaging over space and time.

%Finally we note that the range of $m$ giving the scaling exponents depends on the number of data of each set
%\footnote{
%For a given set of data with a gaussian distribution,
%the calculation of $S_m(\tau)$ from (\ref{scaling}) depends on the number of data.
%Let us consider two types of calculation, one (i) for an infinite number of data, the other (ii) for a number of data equal to the one in (M), (G) or (D). For a given order $m$, we can calculate the standard deviation error between  $S_m^{(i)}(\tau)$ and $S_m^{(ii)}(\tau)$. We find that it increases with $m$. The maximum value of $m$
%reported in figure \ref{GSF} bottom-right for each set (M), (G) or (D), corresponds to a standard deviation lower than 20 \%.}.

The exponents can be fitted to the standard $p$-model derived for hydrodynamic \citep{Meneveau87} and magnetohydrodynamic turbulence \citep{Carbone93}.
This model is defined by
\begin{equation}
\zeta(m)=1-\log_2\left[p^{m/3} + (1-p)^{m/3}\right].
\end{equation}
We find $p_{\rm (G)}=0.68$, $p_{\rm (M)}=0.68$, and $p_{\rm (D)}=0.83$.
The last value compares surprising well with those for the solar wind measured by the Ulysses spacecraft \citep{Pagel02,Nicol08},
%grs
%and for the magnetospheric cusp measured by the Polar satellite \cite{Yordanova04}, even if the frequency ranges
and for the magnetospheric cusp measured by the Polar satellite \citep{Yordanova04}, even though the frequencies
%grs
%are several orders of magnitude different.
differ by several orders of magnitude.

%grs
%In conclusion, the wavelet spectral analysis of sunspot records has revealed two different behaviors, depending whether the solar magnetic cycle is at its minimum or maximum. This suggests two different kinds of turbulence in the convection zone, driven either by inertia or rotation. The signature of such a fully developed turbulence is confirmed by the calculation of the GSF scaling exponents which indicate strong intermittency.
\section{Summary}
In conclusion, the wavelet spectral analysis of sunspot records has revealed two different behaviors, depending on whether the
Sun is quiescent or active.
%solar magnetic cycle is at its minimum or maximum.
This suggests two different kinds of turbulence in the convection zone, controlled either by inertia or by rotation. The signature of such fully developed turbulence is confirmed by the calculation of the GSF scaling exponents, which indicate strong intermittency.

\section*{Acknowledgments}
We are grateful for support from the Dynamo Program at KITP (supported in part by the National Science Foundation under Grant No. PHY05-51164), during which this work was started. We thank Prof. Steve Tobias for helpful comments, and Prof. Ilya Usoskin for providing the $^{10}$Be data. Finally F.P. and R.S. are grateful for support from a RFBR/CNRS 07-01-92160 PICS grant.

\bibliographystyle{mn2e} % this style correspond to journal style
\bibliography{ref}

\label{lastpage}
\end{document}